\documentclass[aps, prx, amsmath, amssymb, reprint,floats,floatfix]{revtex4-2}
\usepackage{graphicx}
\usepackage{bm}
\usepackage[utf8]{inputenc}
\usepackage[T1]{fontenc}
\usepackage{braket}
\begin{document}

\title{Three-fold way of entanglement dynamics in monitored quantum circuits}

\author{T. Kalsi}
\email{t.kalsi@lancaster.ac.uk}
\author{A. Romito}
\author{H. Schomerus}
\affiliation{Department of Physics, Lancaster University, Lancaster LA1 4YB, United Kingdom}
\date{\today}

\begin{abstract}
We investigate the measurement-induced entanglement transition in quantum circuits built upon Dyson's three circular ensembles (circular unitary, orthogonal, and symplectic ensembles; CUE, COE and CSE). We utilise the established model of a one-dimensional circuit evolving under alternating local random unitary gates and projective measurements performed with tunable rate, which for gates drawn from the CUE is known to display a transition from extensive to intensive entanglement scaling as the measurement rate is increased. By contrasting this case to the COE and CSE, we obtain insights into the interplay between the local entanglement generation by the gates and the entanglement reduction by the measurements. For this, we combine exact analytical random-matrix results for the entanglement generated by the individual gates in the different ensembles, and numerical results for the complete quantum circuit. These considerations include an efficient rephrasing of the statistical entangling power in terms of a characteristic entanglement matrix capturing the essence of Cartan's KAK decomposition, and a general result for the eigenvalue statistics of antisymmetric matrices associated with the CSE.
\end{abstract}
\maketitle

\section{Introduction}
In many-body quantum systems, interactions between particles create quantum correlations, termed entanglement.
By virtue of these correlations, the information encoded by quantum systems quickly becomes inaccessible if the associated degrees of freedom couple with the environment. This effective information loss can be captured by quantum statistical methods, such as the von Neumann entanglement entropy, whose scaling properties with system size can be used to characterise ergodic thermal-like or nonergodic localised phases \cite{Znidaric2008Many-bodyField, Bauer2013AreaOrder, Kjall2014Many-BodyChain, Luitz2015Many-bodyChain, Chan2019Unitary-projectiveDynamics, Eisert2010Colloquium:Entropy, Bardarson2012UnboundedLocalization, Szyniszewski2019EntanglementMeasurements, Goto2020}.
Open quantum systems in thermal equilibrium display an extensive scaling of entanglement entropy between the system and its environment, often referred to as a \textit{volume law}. Thermalisation erases local memory of the initial system conditions and configuration, due to the coupling of the quantum system to an external environment (‘reservoir’) with which the system can exchange energy. An extensive scaling of entanglement entropy can also occur in a certain class of closed systems, as detailed by the Eigenstate Thermalisation Hypothesis (ETH) \cite{Deutsch1991QuantumSystem, Srednicki1994ChaosThermalization, DAlessio2016FromThermodynamics, Borgonovi2016QuantumParticles, RevModPhys.91.021001, Nandkishore2015Many-bodyMechanics}. This results in the attainment of an effective thermal equilibrium in a finite time, manifesting ergodic dynamics and internally highly entangled states.
However, when the interactions are constrained to be local, the associated entanglement entropy may be restricted to a sub-extensive \textit{area-law} scaling, violating the ETH and suggesting a \textit{volume-to-area-law} entanglement transition.

Presently, we know of two key paradigms of this entanglement transition.  The first is many-body localisation, which arises in locally interacting isolated systems that are furthermore strongly disordered \cite{Znidaric2008Many-bodyField, Basko2006Metal-insulatorStates, Gornyi2005InteractingTransport, Bauer2013AreaOrder, Kjall2014Many-BodyChain, Luitz2015Many-bodyChain, Altman2015UniversalSystems, Nandkishore2015Many-bodyMechanics, Geraedts2016, Bardarson2012, https://doi.org/10.1002/andp.201600350, PhysRevB.82.174411}.
The second, which provides the backdrop for the present work, are randomly driven many-body systems monitored by their environment, as modelled by a one-dimensional (1D) quantum circuit with random local interactions and fully recorded measurements
\cite{Chan2019Unitary-projectiveDynamics, Skinner2019Measurement-InducedEntanglement,  Li2018QuantumTransition, Li2019Measurement-drivenCircuits,  Szyniszewski2019EntanglementMeasurements, Li2021, Gullans2019purification, Gullans2020, Zabalo2020CriticalCircuits, PhysRevB.104.155111, Zabalo2021, Li2020Conformal, Bao2020, Jian2020, Fan2021, Bao2021, Bera2020, Sang2021measurement, Zhang2020, Choi2020, Nahum2021, Rossini2020, Iaconis2020, Szyniszewski2020UniversalityEntanglement}.
The unobserved circuit is characterised by an entanglement entropy that grows ballistically \cite{Nahum2017QuantumDynamics, vonKeyserlingk2018OperatorLaws, Nahum2018OperatorCircuits}
until saturation in a quasistationary regime
obeying a volume law (hence, a linear scaling of the quasistationary entropy with system size). Local measurements with recorded outcomes keep the system in a pure state, but induce stochastic nonunitary backaction counteracting this growth. These hybrid unitary-projective circuits have been used to establish that sufficiently frequent local projective measurements drive a transition to a sub-extensive area-law scaling of the quasistationary entanglement entropy, which in this 1D setting hence becomes independent of the system size. 
Indeed,  local measurements performed with sufficient frequency and strength manifest the quantum Zeno effect \cite{Misra1977TheTheory, doi:10.1119/1.12204, PhysRevResearch.2.033512} in which dynamics are frozen and entropy does not grow at all.

From a more general perspective, much of our understanding of complex quantum dynamics is built on the identification of universality classes based on symmetries of the system \cite{Haake2018QSoC}. This is firmly embedded in random-matrix theory (RMT), where Dyson's three-fold way in terms of time-reversal symmetry \cite{Dyson1962StatisticalI, Mezzadri2007HowGroups} paved the way to a comprehensive classification that now also includes topological aspects \cite{doi:10.1063/1.3149495, Ryu_2010}, and furthermore started to reach beyond the case of unitary dynamics \cite{RevModPhys.93.015005}.
In the original design of the above-mentioned circuits \cite{Skinner2019Measurement-InducedEntanglement, Li2019Measurement-drivenCircuits, Chan2019Unitary-projectiveDynamics, Li2018QuantumTransition}, the unitary evolution is implemented via two-site unitary matrices $U$ drawn from the circular unitary ensemble (CUE), corresponding to systems in which time-reversal symmetry is broken.
Circuits in which the gates are sampled from the Clifford group
render numerics computationally tractable for large system sizes \cite{Li2019Measurement-drivenCircuits, Zabalo2020CriticalCircuits, PhysRevB.104.155111, PhysRevLett.127.235701, Turkeshi2020, Gullans2020, Gullans2019purification, Li2021, Zabalo2021}. Other circuit variations capture special dynamical cases, for instance, deterministic unitary evolution preserving Gaussian states \cite{Cao2019, Chen2020, Jian2020tn, Alberton2021, Buchhold2021, Tang2021, Zhang2021, kells2021topological, altland2021dynamics}, and dual-unitary circuits in which unitary transfer matrices describe evolution both in time and space, protecting the ballistic spread of entanglement \cite{Bertini2019ExactDimensions, PhysRevB.101.094304, doi:10.1063/5.0056970}. This dual-unitary class of circuits includes instances of both integrable (i.e. with local conservation laws) and chaotic unitary dynamics. Furthermore, quantum circuits based on an array of qudits (of local Hilbert space dimension \textit{q}) are generic models of chaotic quantum dynamics \cite{Nahum2017QuantumDynamics, Rakovszky2018DiffusiveConservation, vonKeyserlingk2018OperatorLaws, Khemani2018OperatorLaws, Chan2018SolutionChaos, Chang2019EvolutionDynamics}.
The circuits can also be adapted to display quasi-continuous stochastic dynamics, in which the entanglement transition can be induced via continuous \cite{Szyniszewski2019EntanglementMeasurements} or variable-strength measurements \cite{Szyniszewski2020UniversalityEntanglement}.

In this work, we aim to obtain further insights into the entanglement dynamics in these systems by going back to the original quantum-circuit design, and contrasting the behaviour with gates drawn from Dyson's three unitary ensembles. Besides the CUE mentioned above, this therefore also encompasses the circular orthogonal ensemble (COE) for systems obeying a conventional time-reversal symmetry, and the circular symplectic ensemble (CSE) representing systems with spin-rotation symmetry \cite{Dyson1962StatisticalI,Haake2018QSoC,Mezzadri2007HowGroups}.
To analytically quantify the difference between these circuits on the local level, we utilise a statistical measure of entangling power, which we formulate via a characteristic entanglement matrix that captures the essence of the Cartan KAK decomposition of a two-qubit gate. This characteristic matrix inherits its statistical behaviour from the circular ensembles, but combines them in a unique way. Indeed, a corollary of our work is the derivation of the joint eigenvalue distribution of antisymmetric matrices associated with the CSE, whose square dictates the entangling power in this class. The statistical entangling power then turns out to be the largest in the CUE, followed by the COE, and finally the CSE. It therefore  displays a nonmonotonous dependence on the standard RMT symmetry index $\beta=2,1,4$ for these three different cases \cite{Haake2018QSoC,Beenakker1997RMT}. On the level of the complete circuit, we then show by extensive numerical results that this leads to an entanglement transition also in the COE and CSE, which is shifted towards smaller values of the tunable measurement rate. This is consistent with interpreting the CUE as providing an upper bound for the entanglement production in the other two cases.

This work is organised as follows.
In Section \ref{sec:Analytics}, we introduce the characteristic entanglement matrix, and relate its statistical properties to the matrix combinations from the different ensembles. This is supplemented by Appendix A, where we derive the  eigenvalue statistics of antisymmetric matrices associated with the CSE.
In Section  \ref{sec:Model}, we describe the quantum circuit model and its numerical implementation, and present the results on the entanglement transition.
We collect our conclusions in Section \ref{sec:Conclusions}.

\section{\label{sec:Analytics} Local entanglement generation}
Quantum circuits consist of many degrees of freedom evolving in discrete time steps under local gate operations, manifesting complex dynamics in which entanglement builds up over time. To capture the essence of these dynamics in the different universality classes, we first obtain analytical results for the statistics of the local entanglement generation, which is the focus of this section.
This involves gates acting on pairs of qubits, corresponding to 4$\times 4$-dimensional unitary matrices.
We proceed in two steps. First, we describe the statistics of entanglement generated by a given unitary matrix when acting onto separable states. This can be quantified by a suitable entangling power \cite{Zanardi2000EntanglingEvolutions,PhysRevA.63.062309,PhysRevA.87.022111}, which we here formulate efficiently by introducing a characteristic entanglement matrix, and then further evaluate with focus on the average  entanglement generated by the gate.
In the second step, we then analyse the complete statistics of this entangling power when the gates are drawn randomly from the appropriate circular ensembles.

\subsection{ Statistics of entanglement generated by a fixed gate}

All transformations $U \in \mathrm{SU(4)}$  on the two-qubit Hilbert space of states
\begin{equation}
\ket{\psi}=\alpha\ket{00}+\beta\ket{01}+\gamma\ket{10}+\delta\ket{11}
\end{equation}
are either local or non-local, categorised by how they act on the components of the bipartite system. Local operations act on one component of the bipartite system, thus can be expressed as a tensor product of one-qubit operations $A_i \in \mathrm{SU(2)}$,
\begin{equation}
    U_\mathrm{local} = A_1 \otimes A_2 \in \mathrm{SU(2)} \otimes \mathrm{SU(2)}.
\end{equation}
As such, local operations cannot change the entanglement properties of the bipartite system. This is different for non-local two-qubit gates, which cannot  be written in this form and do change the entanglement properties of the state. The capacity of these gates to produce entanglement in a bipartite system can be quantified via a suitably defined \emph{entangling power}, typically utilising the entanglement entropy post-operation maximised or averaged over all separable states \cite{PhysRevA.63.062309,PhysRevA.87.022111,Zanardi2000EntanglingEvolutions, Morachis2021EntanglingGates}.
Here, we implement this notion via a characteristic entanglement matrix, to which we are guided by studying the average of the squared concurrence
\begin{equation}
\mathcal{C}^2=4|\alpha\delta-\beta\gamma|^2
\end{equation}
in the normalised post-operational state.

Let us therefore consider the  effect of acting by an arbitrary but fixed two-qubit gate $U$ on a random initial separable state $\ket{\phi} = \ket{\chi_1} \otimes \ket{\chi_2}$. In the computational basis,
this state can be written as
\begin{align}
\ket{\phi}&=(a\ket{0}+b\ket{1})\otimes(c\ket{0}+d\ket{1})
\nonumber
\\
&=ac\ket{00}+ad\ket{01}+bc\ket{10}+bd\ket{11},
\end{align}
so that the squared concurrence vanishes.
For the post-operation state $\ket{\psi}=U\ket{\phi}$, however, we obtain a finite result, which we average by assuming independent uniform distributions of the two input states $\ket{\chi_i}$ over their respective Bloch spheres. This result can be written compactly as
\begin{align}
    \overline{\mathcal{C}^2} &= \frac{4}{9}- \frac{1}{36} |\mathrm{tr}\,V|^2,
    \label{eq:cv}
\end{align}
where
\begin{align}
    V&=Y U^T Y U, \quad  Y = \sigma_y \otimes \sigma_y
        \label{eq:vmat}
\end{align}
is a characteristic matrix that efficiently captures the essence of the intrinsic entanglement characteristics.

To demonstrate this, we make use of Cartan's KAK decomposition \cite{Tucci2005AnProgrammers}, which allows us to  explicitly re-express any two-qubit operation via unitary single-qubit operations and 3 entanglement parameters. Explicitly, the decomposition asserts that for any $ U \in \mathrm{SU}(4)$, there exist single-qubit gates $A_i, B_i \in \mathrm{SU}(2)$ and parameters $a,b,c \in \mathbb{R}$ such that
\begin{equation}
    U = (A_1 \otimes A_2) \exp[i(a,b,c)\cdot(X,Y,Z)] (B_1 \otimes B_2),
\end{equation}
where $(X,Y,Z)$ is the vector of the tensor products of the single-qubit Pauli matrices, $X=\sigma_x \otimes \sigma_x$ and $Z=\sigma_z \otimes \sigma_z$, analogous to the definition of $Y$ above. Hence, KAK parameterises the 15-parameter Lie Group {SU(4)} such that 12 parameters (those associated with {SU(2)}) characterise local operations, and 3 parameters ($a,b,$ and $c$) characterise non-local operations, and does this for all $U \in \mathrm{SU}(4)$ (including all matrices in the three circular ensembles).
Noting that $v_n^*=\sigma_y v_n\sigma_y$ for any $\mathrm{SU}(2)$ matrix $v_n$ enables us to re-express the characteristic matrix $V$ as
\begin{equation}
V =  (B_1 \otimes B_2)^\dagger \exp[2i(a,b,c)\cdot(X,Y,Z)] (B_1 \otimes B_2).
\end{equation}
That is, the characteristic matrix $V$ and the matrix exponential $\exp[2i(a,b,c)\cdot(X,Y,Z)]$ are related by a basis change, such that their eigenvalues coincide,
\begin{align}
\lambda_1&\equiv\exp(i\varphi_1)=\exp(-2i(a+b+c)),
\\
\lambda_2&\equiv\exp(i\varphi_2)=\exp(2i(-a+b+c)),
\\
\lambda_3&\equiv\exp(i\varphi_3)=\exp(2i(a-b+c)),
\\
\lambda_4&\equiv\exp(i\varphi_4)=\exp(2i(a+b-c)),
\end{align}
depending only on the parameters $a$, $b$ and $c$,
\begin{align}
a&=(\varphi_3+\varphi_4-\varphi_1-\varphi_2)/8,
\\
b&=(\varphi_2+\varphi_4-\varphi_1-\varphi_3)/8,
\\
c&=(\varphi_2+\varphi_3-\varphi_1-\varphi_4)/8.
\end{align}
Different permutations of the eigenvalues $\varphi_i$ deliver combinations of $a, b,$ and $c$ that differ only by the single-qubit operations. The parameters $a$, $b$ and $c$ are furthermore independent of an overall $\mathrm{U}(1)$ phase in the matrix $V$. While analogous relations have been derived before \cite{PhysRevA.87.022111}, they usually rely on the isomorphism from $\mathrm{SU}(2)\times\mathrm{SU}(2)$ to $\mathrm{SO}(4)$. With the characteristic matrix $V$, the parameters $a$, $b$ and $c$ can be obtained directly from the standard  definition of the gate.

Based on these observations, we can then derive explicit expressions for the entanglement characteristics using the data from $V$. This is already manifest in Eq.~\eqref{eq:cv} for the averaged square of the concurrence, which can be further expressed as
\begin{equation}
\overline{\mathcal{C}^2} = \frac{1}{3} - \frac{1}{9}(
   \cos{4a}\cos{4b} + \cos{4a}\cos{4c} + \cos{4b}\cos{4c}),
   \end{equation}
which is completely symmetric in the real parameters $a,b,$ and $c$.
The corresponding averaged purity of the reduced density matrix of each qubit then follows from
\begin{equation}
    \overline{\mathcal{P}} = 1-\overline{\mathcal{C}^2}/2.
\end{equation}
From here on, it is useful to capture this data in the
quantity
\begin{equation}
    t=\frac{1}{16} |\mathrm{tr}(V)|^2, \quad 0\leq t \leq 1.
\end{equation}
The different circular ensembles impose different constraints on the structure of $U$, and hence, the parameters $a,b$ and $c$, translating into distinct statistics of the entanglement characteristic $t$, to which we turn next.

\subsection{\label{sec:RMT} Statistics of entanglement generated by random gates}
Rather than only averaging over all possible initial separable states, we can consider the gate $U$ itself being randomly taken from one of the standard circular ensembles. This translates into the statistics of the characteristic matrix $V$, particularly its eigenvalues, which then determine the statistics of the quantity $t$.

To do this, we exploit the fact that matrices of the CUE can be used to generate matrices from the other circular ensembles \cite{Mezzadri2007HowGroups}. Given $W$ is in the $\mathrm{CUE}$, $W^TW$  is in the $\mathrm{COE}$, and $-JW^TJW$  in in the $\mathrm{CSE}$, where $J = i\sigma_y \otimes \openone_2$. The ensembles of characteristic matrices  to consider are therefore
\begin{equation}
V=
\left\{
  \begin{array}{ll}
    YW^TYW, & \mbox{($U$ in CUE)} \\
    YW^TWYW^TW, & \mbox{($U$ in COE)} \\
    -YW^TJWYW^TJW, & \mbox{($U$ in CSE)}
  \end{array}
\right.
\end{equation}
with $W$ from the CUE. It proves helpful to introduce the matrix $y=(\openone +Y)/\sqrt{2} =y^T$, which satisfies $y^2=Y$ and $y^{-1}=Yy$. Using the measure-preserving substitutions $W\to y W y$ for the CUE, $W\to W y$ in the COE and CSE, and the eigenvalue-preserving similarity transformation $V \to yVYy$, we can then eliminate the $Y$ matrices without affecting the statistics. This  gives the equivalent ensembles
\begin{equation}\label{eq:effectiveensembles}
V=
\left\{
  \begin{array}{lll}
    W^TW, & \mbox{($U$ in CUE)} & \to V  \mbox{ in COE} \\
    (W^TW)^2, & \mbox{($U$ in COE)}  & \to V  \mbox{ in COE$^2$}\\
    -(W^TJW)^2, & \mbox{($U$ in CSE)} & \to V  \mbox{ in A$^2$}
  \end{array}
\right.
\end{equation}
where $\mathrm{A}$ is the ensemble of antisymmetric matrices $A = iW^T J W $ associated with the CSE \cite{Bardarson2008Proof}. For $U$ in the CUE or COE, the eigenvalue statistics of $V$ can be read off directly from the known statistics of the COE, obeyed by the eigenvalues $\varphi_i$ or $\varphi_i/2$. For $U$ in the CSE, we derive the joint eigenvalue distribution for matrices $A^2$ with arbitrary matrix dimension in Appendix \ref{sec:AppendixA}.  This takes the form of a CUE eigenvalue distribution with 2 two-fold degenerate eigenvalues. Applied to the two-qubit gates, this exact eigenvalue degeneracy implies that only one of the three KAK parameters $a,b$ or $c$ is ever finite. We then only need the distribution of the one finite parameter, which we identify with the parameter $a$.

\begin{widetext}
In terms of the parameters $a$, $b$, and $c$, we then obtain the joint distributions
\begin{align}
P(a,b,c)\propto
\left\{
  \begin{array}{ll}
    |\sin(2b+2c)\sin(2a+2c)\sin(2a+2b)\sin(2a-2b)\sin(2a-2c)\sin(2b-2c)|, & \mbox{($U$ in CUE)}\\
     |\sin(b+c)\sin(a+c)\sin(a+b)\sin(a-b)\sin(a-c)\sin(b-c)|, & \mbox{($U$ in COE)}\\
\sin^2(2a).  & \mbox{($U$ in CSE)}
  \end{array}
\right.
\end{align}
\end{widetext}
These joint distributions capture the complete statistics of the entanglement characteristics in each ensemble.
For instance, they imply the ensemble averages of $t$,
\begin{align}
\overline{t}=
\left\{
  \begin{array}{ll}
    1/10, & \mbox{($U$ in CUE)}\\
    23/140, & \mbox{($U$ in COE)}\\
1/4.  & \mbox{($U$ in CSE)}
  \end{array}
\right.
\end{align}
In the CSE, where $t=\cos^2(2a)$, we can furthermore obtain the complete probability distribution
\begin{equation}
P_{CSE}(t)=\frac{2}{\pi}\sqrt{t^{-1}-1}.
\end{equation}
In the entangling power \eqref{eq:cv}, the quantity $t$ appears with a minus sign, such that the averaged squared concurrence $\mathcal{C}^2$ is the largest for the CUE, and the smallest for the CSE. Notably, this means that the entangling power follows an unconventional ordering in the standard RMT symmetry index $\beta$, given by $\beta=1$ for the COE, $\beta=2$ for the CUE, and $\beta=4$ for the CSE, setting this characteristics well apart from the behaviour of other observables \cite{Haake2018QSoC,Beenakker1997RMT}.

This concludes our considerations of the entanglement generation by random gates.
We now turn to question how this affects the entanglement transition in monitored quantum circuits, where these processes compete with measurements that suppress the entanglement.

\section{\label{sec:Model} Entanglement transition}

\begin{figure}[b]
\includegraphics[width=8.5cm]{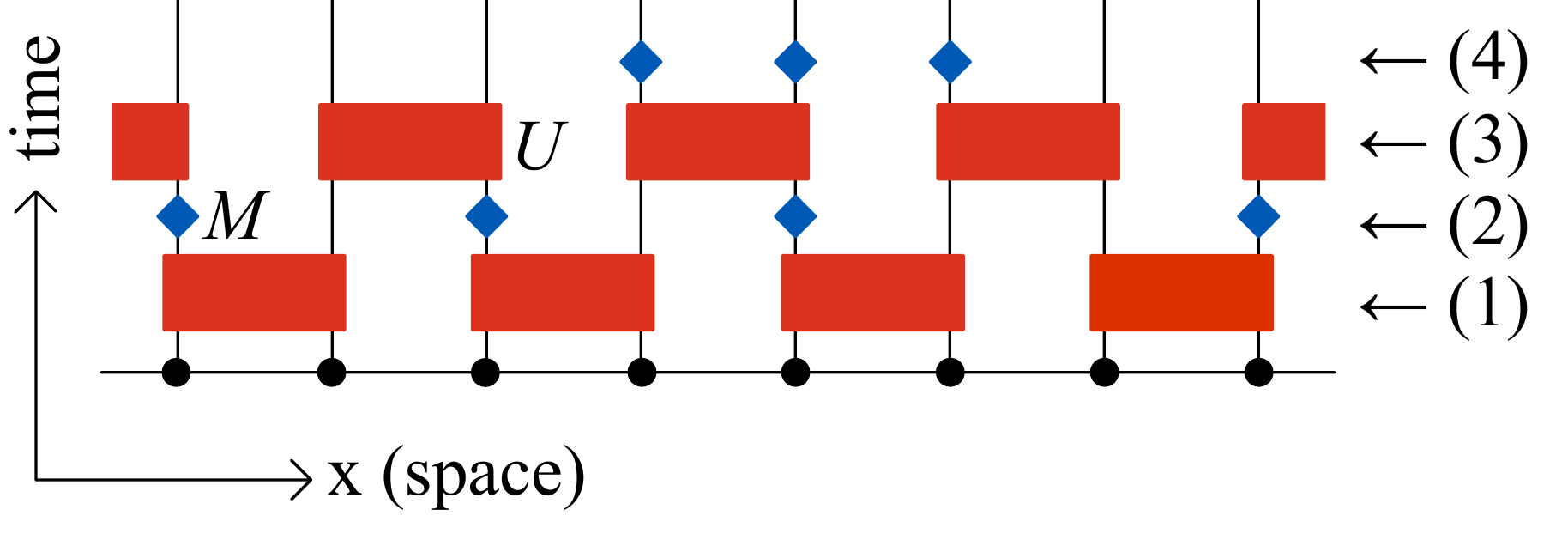}
\caption{\label{fig:model} Schematic illustration of the unitary-projective time evolution of a one-dimensional quantum circuit during a single time step. Each of the $L$ spin locations are denoted by solid black dots. Alternating unitary operations $U$ on pairwise spins are represented by red rectangular blocks and generate entanglement between adjacent degrees of freedom, while blue diamond shapes represent local projective measurements $M$ implemented with probability $p$, which remove entanglement. The illustration depicts one possible random realisation of the measurements.}
\end{figure}

\begin{figure*}[bth]
\centering
\includegraphics[width=0.9\linewidth]{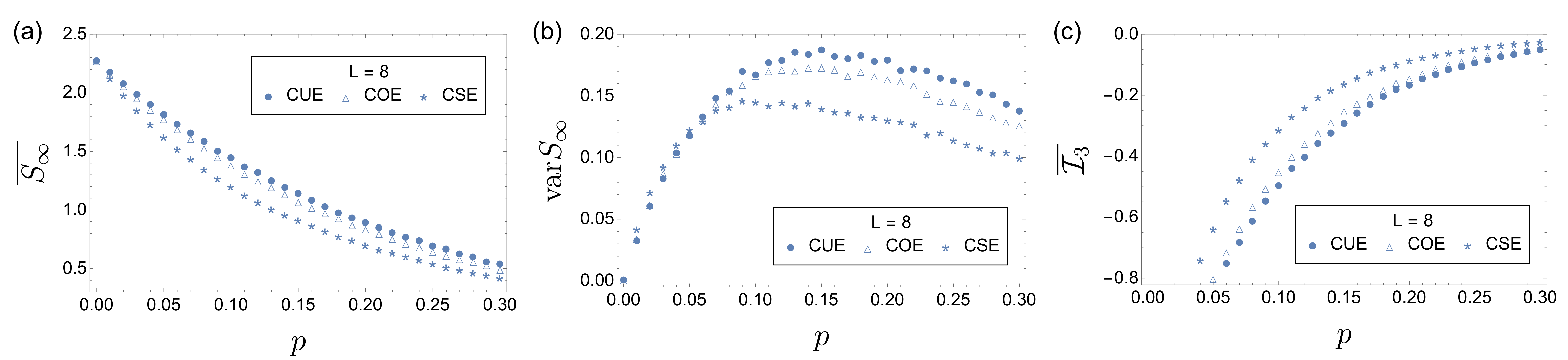}
  \caption{\label{fig:L8} Comparison of the quasistationary entanglement statistics in quantum circuits of size $L=8$, with the unitary dynamics generated from three different universality classes, {CUE} (filled circles), {COE} (open triangles), and {CSE} (stars). (a) Ensemble-averaged quasistationary entanglement entropy $\overline{S_\infty}$, (b) corresponding fluctuations $\mathrm{var}S_\infty$, and (c) tripartite mutual information $\overline{\mathcal{I}_3}$, all plotted as a function of the measurement rate $p$. Each data point is obtained from  $10^5$ realisations.}
\end{figure*}

To determine the role of the gates for the entanglement statistics in larger systems,
we first describe the model of a hybrid unitary-projective quantum circuit, and then present numerical results for the volume-to-area-law entanglement transition in systems based on the three different ensembles.
\subsection{Model}
The numerical model implemented is based on the universal quantum-circuit architecture \cite{Chan2019Unitary-projectiveDynamics, Nahum2017QuantumDynamics, vonKeyserlingk2018OperatorLaws, Nahum2018OperatorCircuits, Szyniszewski2019EntanglementMeasurements} diagrammatically shown in Fig.~\ref{fig:model}. The quantum circuit is a spatially periodic
one-dimensional chain of $L$ spins, with $L$ taken to be even. The quantum state of the chain is evolved in discrete time steps, each of which consists of  four layers of operations. The first layer applies unitary two-spin operators $U$ to consecutive pairs of spins, namely, the spins labelled by indices $2l-1$ and their next-neighbouring site $2l$ ($l=1,\ldots,L/2$).
The second layer implements projective measurements $M$ of the $z$-component of each spin, which for each spin are carried out independently with a probability $p$. The third operation applies unitary two-spin operators $U$ to each even and next-neighbouring site, labelled by indices $2l$ and $2l+1 \mod L$. The fourth and final layer again implements independent projective measurements $M$ on each spin, with the same probability $p$ as in the second layer.

The variation of unitary operations and measurements throughout space and time manifests a stochastic time evolution. In particular, the local measurements induce stochastic nonunitary backaction, reducing quantum correlations---indeed, projecting the system onto an eigenstate of a local operator completely disentangles the corresponding local degree of freedom from the rest of the system. In the quantum circuit, the measurements are implemented by the standard von Neumann
protocol \cite{vonNeumann1938MathematicalTheory, Wiseman2009QuantumControl, Jacobs2014QuantumApplications}, for which measurement of an observable with the Hermitian operator $\hat{O} = \sum_{i,n} \lambda_i \ket{i,n} \bra{i,n}$ in state $\ket{\psi}$ results in real outcomes restricted to the eigenvalues $\lambda_i$. These occur with probability $p_i= \bra{\psi}P_i\ket{\psi}$, where $P_i = \sum_n\ket{i,n}\bra{i,n}$ is the projector onto the associate subspace of eigenstates $\ket{i,n}$, taken here to be suitably orthonormalised. This projector also determines the measurement backaction,
\begin{equation}
\ket{\psi}
= \frac{P_i}{\sqrt{p_i}} \ket{\psi}.
\end{equation}
Importantly, it is assumed that all measurement results are recorded, so that the system remains in a pure state.

In the original circuit design \cite{Skinner2019Measurement-InducedEntanglement, Li2019Measurement-drivenCircuits, Chan2019Unitary-projectiveDynamics, Li2018QuantumTransition}, these measurements were combined with  unitary two-spin operators that  were drawn independently from the circular unitary ensemble (CUE), each generating entanglement locally as considered in the previous section. In the dynamics of the quantum circuit, the entanglement then spreads out over the system, until the system settles into a quasistationary state whose entanglement characteristics depend on the measurement rate $p$. It then was found that the system displays a transition in the von Neumann
bipartite entanglement entropy
\begin{equation}
   S_A = - \mathrm{tr}(\rho_A \ln \rho_A ),
\end{equation}
where $\rho^A$ is the reduced density matrix of subchain $A$ with length $L/2$, calculated by tracing out the subsystem complementary to $A$. In this transition, $S_A$ scales linearly with system size $L$ below a critical measurement rate $p_c$, manifesting a volume law, but becomes independent of system size for $p>p_c$, manifesting an area law.
In the remainder of this work, this transition serves as our reference point for numerical investigations, where we  extend these considerations to the case of  gates obtained from the different circular ensembles.

\subsection{\label{sec:Numerics}Numerical results}

Our goal is to establish the persistence of a measurement-induced entanglement transition in the model in which entanglement dynamics are now generated by either the COE or CSE, and to compare the results with the CUE.
In each case, the unitary two-spin operators $U$ are therefore generated randomly and independently with probability distribution given by the corresponding Haar measure \cite{Mezzadri2007HowGroups, Anderson2009AnMatrices}, while the measurement protocol is kept as before.
We consider the dynamics of a system of length $L$ whose initial state is separable, thus characterised by a vanishing entanglement entropy. We then obtain the numerically averaged bipartite von-Neumann entropy $\overline{S}(t)$ as a function of the discrete iteration time $t$, where the average is carried out over different realisations of the unitary gates and randomly applied measurements and outcomes, as well as the two inequivalent choices of bipartition (cutting through a pair of spins entangled in the last layer, or between such pairs).
In the quasistationary regime, here always fully attained after $4L$ time steps, the steady-state entropy $\overline{S_\infty}$ is consistently observed to be independent of the initial state, so that we can set it to $\ket{\psi(t=0)}=\ket{000...0}$.

\begin{figure*}[t]
\centering
 \includegraphics[width=0.9\linewidth]{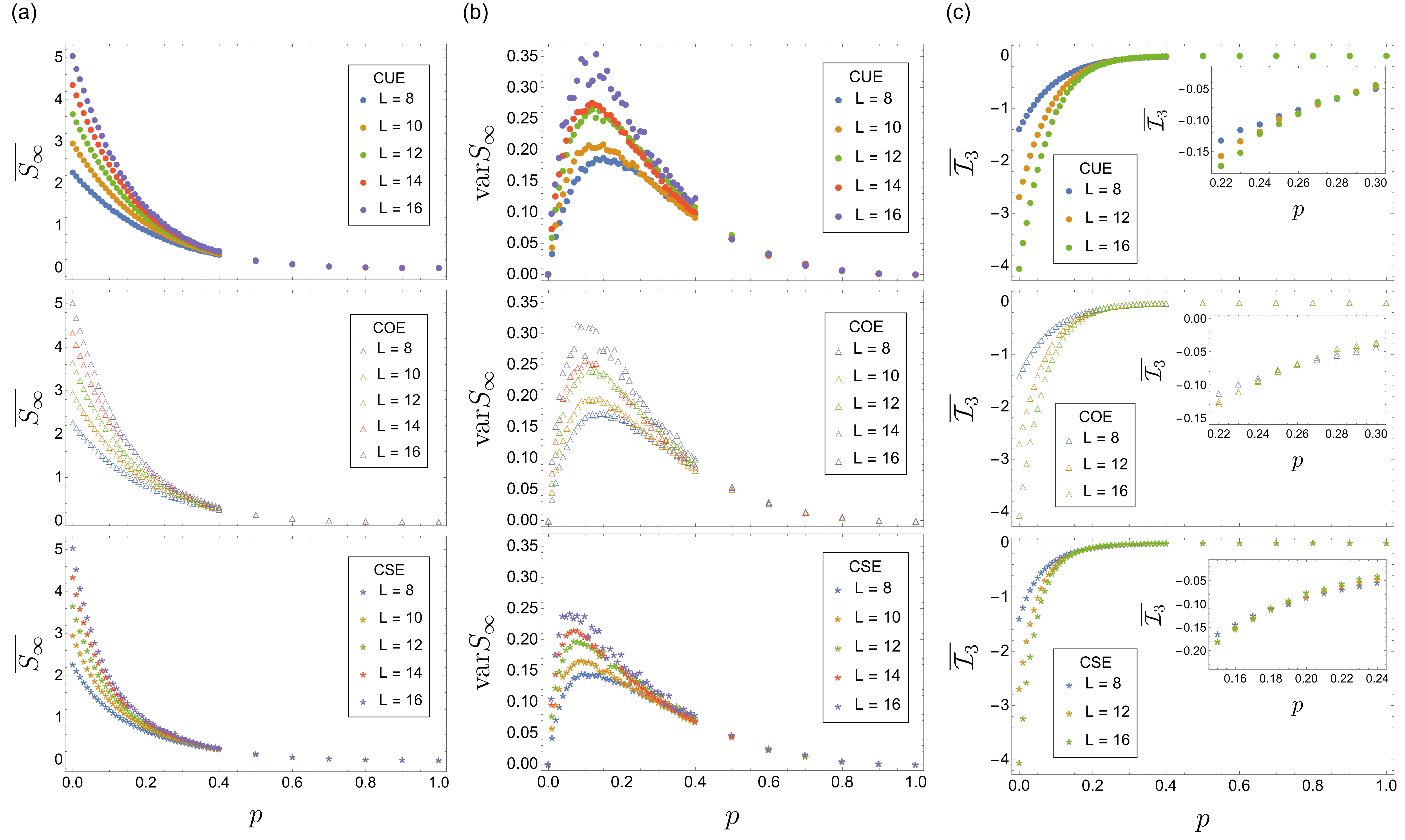}
  \caption{\label{fig:numerics} Quasistationary entanglement statistics  as in Fig.~\ref{fig:L8}, but now comparing systems of size $8\leq L\leq 16$ in each ensemble, {CUE} (top row), {COE} (middle row), and {CSE} (bottom row). The insets in (c) focus on the region where tripartite mutual information crosses for different system sizes. Each data point is obtained from $10^5$ realisations ($10^4$ for $L=16$).}
\end{figure*}

Figure \ref{fig:L8} depicts the resulting entanglement statistics generated by the three circular ensembles for quantum circuits of fixed size $L=8$,
as a function of the projective measurement rate $p$.
Panel (a) shows the steady-state entanglement entropy $\overline{S_\infty}$.
In the purely random unitary evolution (i.e. in the absence of measurements, $p=0$), this entropy consistently recovers the Page result \cite{Page1993AverageSubsystem},
\begin{equation}
    \overline{S_{\infty}} = \frac{L}{2} \ln 2 - \frac{1}{2}.
    \label{eq:page}
\end{equation}
This is a signature of a maximally ergodic system obeying a volume law, $\overline{S_{\infty}} \propto L$, where the state of the system is completely random in Hilbert space.
As the projective measurement rate $p$ increases, the quasistationary entanglement entropy is suppressed, which is a consequence of the measurement backaction.
In comparing the results for the different ensembles, we find that for all parameters, $\overline{S_\infty}$ is systematically the largest in the CUE, and the smallest in the CSE. This is consistent with the differences in the entangling power of individual gates obtained in our analytical considerations, and hence retains the unconventional ordering by the symmetry index $\beta$.

Figure~\ref{fig:L8} (b) verifies these trends by the sample-to-sample fluctuations $\mathrm{var}\ S_\infty$.
Compared to the CUE, these fluctuations are suppressed in the other two ensembles, and most noticeably so in the CSE.
Figure~\ref{fig:L8} (c) substantiates this further in terms of the tripartite mutual information, defined for the three adjacent subsystems \textit{A, B,} and \textit{C}, each of size $L/4$,
\begin{equation}
\begin{aligned}
    \mathcal{I}_3 (A:B:C) &= S_A + S_B + S_C + S_{A \cup B \cup C} \\ &\quad - S_{A\cup B} - S_{A \cup C} - S_{B\cup C}.
\end{aligned}
\label{eq:i3def}
\end{equation}
We again find a systematic shift between the different ensembles, consistent with the quantitative differences in the entanglement generation processes.

At any fixed $p$, the dependence of the entanglement statistics on $L$ further distinguishes an extensive volume-law scaling from an area-law scaling. In Fig.~\ref{fig:numerics}, we analyse this for the different ensembles. Panel (a) plots the steady-state entanglement entropy $\overline{S_\infty}$ as a function of the projective measurement rate $p$, where we now compare results for different system sizes $L$.
In absence of measurements ($p=0$), we continue to consistently recover the Page result \eqref{eq:page}. As long as the projective measurement rate $p$ remains small, the quasistationary entanglement entropy remains large,   and continues to increase linearly with system size $L$.
Increasing the measurement rate further suppresses $\overline{S_\infty}$ monotonically, until its behaviour becomes indicative of being independent of system-size at sufficiently large measurement rates.
This translates to an entanglement transition occurring at a characteristic critical measurement rate $p_c$ for each of the ensembles. As we show next, when compared to the CUE, the value of $p_c$  is slightly reduced in the COE, and noticeably reduced in the CSE.
For this, we first turn in
Fig.~\ref{fig:numerics} (b) to the sample-to-sample fluctuations $\mathrm{var}\ S_\infty$, which become large in the vicinity of the entanglement transition \cite{Szyniszewski2019EntanglementMeasurements}.
Compared to the CUE, these fluctuations are not only suppressed in the other two ensembles,  but the location of maximal fluctuations is also shifted to smaller measurement rates, most noticeably so in the CSE.

Finally, in Fig.~\ref{fig:numerics} (c) we quantify this shift of the transition in terms of the ensemble-averaged tripartite mutual information  $\overline{\mathcal{I}_3}$.
In this quantity, the transition is indicated by the crossing point between curves produced by different system sizes, which must necessarily have length $L$ divisible by 4. Indeed,  $\overline{\mathcal{I}_3}$ can be interpreted akin to an order parameter; in the limit of no measurement ($p=0$), $\overline{\mathcal{I}_3}$ diverges to  infinitely negative values for increasing system size, while it is expected to be a negative number close to zero in magnitude once the area-law phase has been established by frequent projective measurement.
In the original hybrid unitary-projective circuits that serve as our reference, the crossing points between curves $\overline{\mathcal{I}_3}$ of different system size are found to have minimal finite-size drifts, which enables the reliable determination of the critical measurement rate with minimal scaling assumptions. (Conversely, for continuous-time models, crossings are found to drift with increasing system size \cite{Zabalo2020CriticalCircuits, Szyniszewski2019EntanglementMeasurements, Gullans2020DynamicalMeasurements, Kitaev2006TopologicalEntropy, Levin2006DetectingFunction}.)

These features are indeed born out by our data in Fig.~\ref{fig:numerics}(c).
In each ensemble, the curves for different system sizes cross at a characteristic location, which then gives an estimate for the critical measurement rate. These estimates are given by $p_c\approx 0.27$ in the CUE, consistent with the literature \cite{Skinner2019Measurement-InducedEntanglement}, as well as $p_c\approx 0.25$ in the COE, and $p_c\approx 0.18$ in the CSE. The ordering of these values therefore conforms once more to our general picture that statistically, gates from the CUE generate the largest amount of entanglement, and those from the CSE generate the smallest amount.

\section{\label{sec:Conclusions}Conclusions}
In summary, we applied random-matrix theory to analytically characterise the entanglement generated by unitary matrices taken from Dyson's three circular ensembles, representing gates with different behaviour under time-reversal symmetry, and supplemented this by numerical investigations of the ensuing entanglement dynamics in hybrid unitary-projective quantum circuits built from such gates.
Utilising a characteristic entanglement matrix,
we found that gates from the circular unitary ensemble, which are conventionally invoked in these circuits, generate more local entanglement than gates from the circular orthogonal or circular symplectic ensemble, resulting in an unusual ordering of this characteristic in terms of the standard symmetry index $\beta$.
This is reflected by the entanglement dynamics in the circuits, where the gates act
locally to generate entanglement and propagate it over the system, while projective measurements counteract these processes and drive the system towards a low-entropy phase.
This results in a transition from an extensive volume-law scaling to a subextensive area-law scaling, which we found to
persist for dynamics generated by the COE and CSE, but occur at a reduced critical rate of projective measurements.
These results imply that ensuring \emph{local} time-reversal symmetry on the level of individual gates can help to inhibit thermalisation in noisy settings, which is desirable in the storage and manipulation of quantum information.

The considerations in this work can be extended into a number of natural and useful directions.
On the level of the individual gates, this includes the consideration of ensembles from other unitary ensembles, in particular those featuring in the complete ten-fold way \cite{Altland1997Nonstandard}. Furthermore, the symmetries could also be imposed globally for finite iterations of the circuit, which implies relations between gates in different layers, and one could also seek to identify analogues of these symmetries for the measurement protocol.
The quantum gates and circuit geometry can also be extended to multi-level systems, including deterministic systems exhibiting the signatures of quantum chaos \cite{Haake2018QSoC} for which random-matrix theory serves as a natural reference point. This may also serve as a natural arena for our general results obtained for the antisymmetric matrices $A$ associated with the symplectic ensemble, which we obtained as a corollary of this work.

\begin{acknowledgments}
We gratefully acknowledge helpful discussions with Jens Bardarson, Carlo Beenakker, and Karol \.Zyczkoswski.
\end{acknowledgments}

\appendix
\section{\label{sec:AppendixA} Joint eigenvalues distribution in the ensemble $\mathrm{A}^2$}
To obtain the eigenvalue statistics of the characteristic entanglement matrix $V$, Eq.~\eqref{eq:vmat}, for $U$ in the CSE, we
reformulate the ensemble as in Eq.~\eqref{eq:effectiveensembles} and then apply the Brownian motion approach to the matrix $A=iW^TJW$. Consider the stochastic process generated by
\begin{align}
W\to& (\openone-i (\delta t)^{1/2} G - \frac{\delta t}{2} G^2)W(\openone-i (\delta t)^{1/2} H -\frac{\delta t}{2} H^2)
\\
\equiv &W+\delta W,
\end{align}
where $\delta t$ is an infinitesimal parameter and we only keep terms up to $O(\delta t)$. Here $H$ and $G$ are independent random Hermitian matrices from the Gaussian unitary ensemble (GUE), characterised by $\overline{G_{kl}}=\overline{H_{kl}}=\overline{G_{kl}H_{mn}}=0$ and
$ \overline{G_{kl}G_{mn}}=\overline{H_{kl}H_{mn}} =N^{-1}\delta_{kn}\delta_{lm}$.
Some key averages that we need are
\begin{equation}
\overline{HMH}=N^{-1}\mathrm{tr}\,M,\quad\overline{HMH^*}=N^{-1}M^T,
\end{equation}
for any fixed matrix $M$.
These expressions imply $\overline {G^2}=\overline {H^2}=\openone$, while cross-terms $\overline{GMH}=0$, and allow us to set
\begin{equation}
\delta W = -i (\delta t)^{1/2} (GW+WH)-\delta t W,
\end{equation}
as this captures all the finite contributions to averages calculated later on.
This process samples the unitary group with the Haar measure. Furthermore, if $W$ is itself from the Haar measure the process is stationary.

Our reference point is the well-known Brownian motion for the eigenvalues $\lambda_n$ of $W$ itself, which results in the standard CUE joint probability distribution.
In perturbation theory (second order in $(\delta t)^{1/2}$ so that we stay accurate in $O(\delta t)$), the eigenvalues change as
\begin{equation}
\delta\lambda_n=\langle n | \delta W | n\rangle+\sum_{m\neq n}\frac{\langle n | \delta W | m\rangle\langle m | \delta W | n\rangle}{\lambda_n-\lambda_m}.
\end{equation}
On average, with help of the formulas above we then have
\begin{equation}
\overline{\delta\lambda_n}=-\delta t\left(\lambda_n+\frac{2}{N}\sum_{m\neq n}\frac{\lambda_n\lambda_m}{\lambda_n-\lambda_m}\right).
\end{equation}
Furthermore, we obtain the correlator
\begin{equation}
\overline{\delta\lambda_n\delta\lambda_m}= -\delta t\frac{2}{N}\lambda_n^2\delta_{nm}.
\end{equation}
We now introduce these averages as drift and diffusion terms into a Fokker-Planck equation, which tells us how the eigenvalue distribution changes,
\begin{equation}
\frac{\partial }{\partial t}P = -\sum_n\frac{\partial }{\partial \lambda_n}\left(
\frac{\overline{\delta\lambda_n}}{\delta t}-\frac{1}{2}\frac{\partial }{\partial \lambda_n}
\frac{\overline{(\delta\lambda_n)^2}}{\delta t} \right)P.
\end{equation}
In the stationary situation, where $W$ is from the CUE Haar measure, the probability distribution must be stationary, too, meaning that the right hand side must vanish.
This is indeed fulfilled, term by term, for the CUE probability distribution
\begin{equation}
P(\{\lambda_n\})\propto \prod_{n<m}|\lambda_n -\lambda_m|^2\sum_k\lambda_k^{-1}.
\end{equation}
(Note that $\lambda_k^{-1}d\lambda_k\propto d\varphi_k$, so that this transforms to the standard eigenphase distribution
$P(\{\varphi_n\})\propto \prod_{n<m}|e^{i\varphi_n} -e^{i\varphi_m}|^2.$)

Now we compare this to the eigenvalue distribution of the matrix $\mathrm{A}^2$ related to the CSE, as defined in Eq.~\eqref{eq:effectiveensembles}.
We will denote the eigenvalues of $A=iW^TJW$ itself as $\mu_n$, and those of $A^2$ as $\lambda_n=\mu_n^2$.
A key point is that for any eigenstate $| n\rangle$,
\begin{equation}
A| n\rangle=\mu_n| n\rangle,
\end{equation}
there is a partner state $| \bar n\rangle$, obtained by taking the complex conjugate of all components, which has the opposite eigenvalue,
\begin{equation}
A| \bar n\rangle=-\mu_n| \bar n\rangle.
\end{equation}
For $A^2$, this implies that $\lambda_n=\lambda_{\bar n}$, and so all these eigenvalues are doubly degenerate.
We also have to take this degeneracy into account in the Brownian motion, which we first formulate for the eigenvalues $\mu_n$.
For instance, in the second-order term we find contributions such as
$\langle n |AH |m \rangle \langle m |AH|n \rangle =\mu_n\mu_m/N$ and
 $\langle n |AH |m \rangle \langle m |H^* A| n \rangle =\mu_n^2\delta_{m\bar n}/N$, where the latter cancels the former for $m=\bar n$. Furthermore, we can combine the contributions of the remaining pairs $m,\bar m \neq n$, and hence only sum over one representative in each pair [we write this as a sum over the pairs $(m,\bar m)$].
Carrying this out in detail gives
\begin{align}
\overline{\delta\mu_n}&=-\delta t\left( 2(1-1/N)\mu_n+\frac{8}{N}
\sum_{\substack{(m,\bar m)\\ \neq (n,\bar n)}}
\frac{\mu_n\mu_m^2}{\mu_n^2-\mu_m^2}\right),
\\
\overline{\delta\mu_n\delta\mu_m}&= -\delta t\frac{4}{N}\mu_n\mu_m(\delta_{nm}+\delta_{\bar nm}).
\end{align}
Next,  we transform this to the corresponding expression of the squared eigenvalues (carefully expanding into second order using Ito calculus),
\begin{align}
\overline{\delta\lambda_n}&=2\overline{\delta\mu_n}\mu_n+\overline{(\delta\mu_n)^2}
\\
&=
-4\delta t\left(\lambda_n+\frac{2}{N/2}
\sum_{m \neq n}
\frac{\lambda_n\lambda_m}{\lambda_n-\lambda_m}\right),
\\
\overline{\delta\lambda_n\delta\lambda_m}&
= 4\mu_n\mu_m\overline{\delta\mu_n\delta\mu_m}
\\
&=
-4\delta t\frac{2}{N/2}\lambda_n^2\delta_{nm}
,
\end{align}
with indices confined to run over distinct eigenvalues only.
We interpret this as the drift and diffusion coefficients in a Brownian motion of $N/2$ numbers, each standing for a pair of eigenvalues, whose joint distribution is again found from the stationarity condition of the corresponding Fokker-Planck equation. But up to an overall factor of 4, the drift and diffusion terms are just the same as in the CUE with $N\to N/2$,
and this overall factor drops out of the stationarity condition.
Therefore, for $U$ in the CSE, the joint probability distribution of the distinct eigenvalues $\lambda_n$ of $V$ coincides with that of CUE matrices with dimension $N/2$. In the main text, we apply this result to the case where $V$ is a $4\times4$ matrix.

%

\end{document}